\documentclass[]{pasj00}
\draft
\usepackage[round]{natbib}
\usepackage{dcolumn}

\begin{document}
\SetRunningHead{M. Serino et al.}{MAXI observations of GRBs}

\title{MAXI observations of GRBs}

\author{
Motoko~\textsc{Serino},\altaffilmark{1}
Takanori~\textsc{Sakamoto},\altaffilmark{2}
Nobuyuki~\textsc{Kawai},\altaffilmark{3,1}
Atsumasa~\textsc{Yoshida},\altaffilmark{2,1}
Masanori~\textsc{Ohno},\altaffilmark{4}
Yuji~\textsc{Ogawa},\altaffilmark{5}
Yasunori~\textsc{Nishimura},\altaffilmark{5}
Kosuke~\textsc{Fukushima},\altaffilmark{6}
Masaya~\textsc{Higa},\altaffilmark{7}
Kazuto~\textsc{Ishikawa},\altaffilmark{3}
Masaki~\textsc{Ishikawa},\altaffilmark{8}
Taiki~\textsc{Kawamuro},\altaffilmark{9}
Masashi~\textsc{Kimura},\altaffilmark{10}
Masaru~\textsc{Matsuoka},\altaffilmark{1,10}
Tatehiro~\textsc{Mihara},\altaffilmark{1}
Mikio~\textsc{Morii},\altaffilmark{1}
Yujin~E.~\textsc{Nakagawa},\altaffilmark{11}
Satoshi~\textsc{Nakahira},\altaffilmark{10}
Motoki~\textsc{Nakajima},\altaffilmark{12}
Yuki~\textsc{Nakano},\altaffilmark{2}
Hitoshi~\textsc{Negoro},\altaffilmark{6}
Takuya~\textsc{Onodera},\altaffilmark{6}
Masayuki~\textsc{Sasaki},\altaffilmark{13}
Megumi~\textsc{Shidatsu},\altaffilmark{9}
Juri~\textsc{Sugimoto},\altaffilmark{1}
Mutsumi~\textsc{Sugizaki},\altaffilmark{1}
Fumitoshi~\textsc{Suwa},\altaffilmark{6}
Kazuhiko~\textsc{Suzuki},\altaffilmark{6}
Yutaro~\textsc{Tachibana},\altaffilmark{3}
Toshihiro~\textsc{Takagi},\altaffilmark{1,6}
Takahiro~\textsc{Toizumi},\altaffilmark{3}
Hiroshi~\textsc{Tomida},\altaffilmark{10}
Yohko~\textsc{Tsuboi},\altaffilmark{7}
Hiroshi~\textsc{Tsunemi},\altaffilmark{13}
Yoshihiro~\textsc{Ueda}\altaffilmark{9}
Shiro~\textsc{Ueno},\altaffilmark{10}
Ryuichi~\textsc{Usui},\altaffilmark{3}
Hisaki~\textsc{Yamada},\altaffilmark{5}
Takayuki~\textsc{Yamamoto},\altaffilmark{1}
Kazutaka~\textsc{Yamaoka},\altaffilmark{14,15}
Makoto~\textsc{Yamauchi},\altaffilmark{5}
Koshiro~\textsc{Yoshidome}\altaffilmark{5}
and
Taketoshi~\textsc{Yoshii}\altaffilmark{3}
} 

\altaffiltext{1}
{MAXI team, Institute of Physical and Chemical Research (RIKEN), 2-1 Hirosawa, Wako, Saitama 351-0198}
\email{motoko@crab.riken.jp}
\altaffiltext{2}
{Department of Physics and Mathematics, Aoyama Gakuin University,\\ 5-10-1 Fuchinobe, Chuo-ku, Sagamihara, Kanagawa 252-5258}
\altaffiltext{3}
{Department of Physics, Tokyo Institute of Technology, 2-12-1 Ookayama, Meguro-ku, Tokyo 152-8551}
\altaffiltext{4}
{Department of Physical Sciences, Hiroshima University, 1-3-1 Kagamiyama, Higashi-Hiroshima, Hiroshima 739-8516}
\altaffiltext{5}
{Department of Applied Physics, University of Miyazaki, 1-1 Gakuen Kibanadai-nishi, Miyazaki, Miyazaki 889-2192}
\altaffiltext{6}
{Department of Physics, Nihon University, 1-8-14 Kanda-Surugadai, Chiyoda-ku, Tokyo 101-8308}
\altaffiltext{7}
{Department of Physics, Chuo University, 1-13-27 Kasuga, Bunkyo-ku, Tokyo 112-8551}
\altaffiltext{8}
{School of Physical Science, Space and Astronautical Science, The graduate University for Advanced Studies, Yoshinodai 3-1-1, Chuo-ku, Sagamihara, Kanagawa 252-5210}
\altaffiltext{9}
{ISS Science Project Office, Institute of Space and Astronautical Science (ISAS), Japan Aerospace Exploration Agency (JAXA), 2-1-1 Sengen, Tsukuba, Ibaraki 305-8505}
\altaffiltext{10}
{Department of Astronomy, Kyoto University, Oiwake-cho, Sakyo-ku, Kyoto 606-8502}
\altaffiltext{11}
{ISS Science Project Office, Institute of Space and Astronautical Science (ISAS), Japan Aerospace Exploration Agency (JAXA), 3-1-1 Yoshinodai, Chuo-ku, Sagamihara, Kanagawa 252-5210}
\altaffiltext{12}
{School of Dentistry at Matsudo, Nihon University, 2-870-1 Sakaecho-nishi, Matsudo, Chiba 101-8308}
\altaffiltext{13}
{Department of Earth and Space Science, Osaka University, 1-1 Machikaneyama, Toyonaka, Osaka 560-0043}
\altaffiltext{14}
{Department of Particle Physics and Astronomy, Nagoya University, Furo-cho, Chikusa-ku, Nagoya, Aichi 464-8601}
\altaffiltext{15}
{Solar-Terrestrial Environment Laboratory, Nagoya University, Furo-cho, Chikusa-ku, Nagoya, Aichi 464-8601}


\maketitle

\begin{abstract}

 Monitor of all-sky image (MAXI) Gas Slit Camera (GSC) 
 detects gamma-ray bursts (GRBs) including the bursts with soft spectra, 
 such as X-ray flashes (XRFs).
 MAXI/GSC 
 is sensitive to
 the energy range from 2 to 30 keV.  
 This energy range is lower than other currently operating instruments 
 which is capable of detecting GRBs.
 Since the beginning of the MAXI operation on August 15, 2009, 
 GSC observed 35 GRBs
 up to the middle of 2013.
 One third of them are also observed by other satellites.
 The rest of them show a trend to have soft spectra and
 low fluxes. 
 Because of the contribution of those XRFs, the MAXI GRB rate is about
 three times higher than those expected 
 from the BATSE log $N$ -- log $P$ distribution.
 When we compare it to the observational results of the Wide-field X-ray 
 Monitor on the High Energy Transient Explorer 2, which covers the 
 the same energy range to that of MAXI/GSC, we find a possibility that 
 many of MAXI bursts are XRFs with $E_{\mathrm{peak}}$ lower than 20 keV.
 We discuss the source of soft GRBs observed only by MAXI.
 The MAXI log $N$ -- log $S$ distribution suggests
 that the MAXI XRFs distribute in closer distance than hard
 GRBs.  
 Since the distributions of the hardness of galactic stellar flares
 and X-ray bursts overlap with those of MAXI GRBs, we discuss
 a possibility of a confusion of those galactic transients with the MAXI 
 GRB samples.
\end{abstract}

\section{Introduction}
 X-ray flashes (XRFs) are subclass of 
 gamma-ray bursts (GRBs) 
 which have significantly softer spectra than those of 
 classical GRBs.
 They are characterized by absence of 
 emission at the high energy band ($>$50 keV)
 \citep{1998ApJ...500..873S,2001grba.conf...16H}.
 Later, the empirical classification of GRBs using fluence ratio in 
 the 2--30 keV
 to 30--400 keV bands was introduced \citep{2005ApJ...629..311S}. 
 While the peak energy values, $E_{\mathrm{peak}}$, in the spectra of
 XRFs distinguish them from classical hard GRBs, 
 \citet{2005ApJ...629..311S} suggest that they arise from the same class
 based on the HETE-2 samples.

 Various models have been proposed to explain the origin of low 
 $E_{\mathrm{peak}}$ in XRFs.
 Some of them do not assume intrinsic difference in the source. 
 Instead, the apparent differences to hard GRBs are caused by 
 the redshift of
 the sources \citep{2001grba.conf...16H} or observer's viewing angles of 
 the GRB jets \citep{2002ApJ...571L..31Y}. Others require the intrinsic
 difference in the conditions of the sources 
 \citep{2002ApJ...578..812M,2002ApJ...581.1236Z,2005ApJ...620..355L}.
 Some of the predictions of those models are investigated against the 
 observed data \citep{2005ApJ...630.1003G,2006A&A...460..653D}.

 Although it is natural to imagine that GRBs which occurred at high 
 redshift have low (apparent) $E_{\mathrm{peak}}$, we have not yet 
 observed such an event.
 So far, all of XRFs with $E_{\mathrm{peak}}$ lower than 20 keV have
 relatively low redshift; the redshift values of
 XRF 020903 \citep[$E_{\mathrm{peak}} < 5$ keV;][]{2004ApJ...602..875S},
 XRF 050416A \citep[$E_{\mathrm{peak}} = 13.67$ keV;][]{2006ApJ...636L..73S},
 and 
 XRF 091018 \citep[$E_{\mathrm{peak}} = 19.43$ keV;][]{2011ApJS..195....2S}
 are 
 0.25 \citep{2004ApJ...606..994S}, 0.6528 \citep{2007ApJ...661..982S},
 and
 0.971 \citep{2009GCN..10038...1C,2012MNRAS.426....2W} 
 respectively.

 There are some attempts to estimate the luminosity function of GRBs 
 \citep{2010ApJ...711..495B,2010MNRAS.406.1944W,
 2010MNRAS.406..558Q,2011MNRAS.417.3025V,2014AAS...22335206L}. 
 However, in order to estimate the luminosity 
 function at high redshift more precisely, it is necessary to observe
 GRBs at lower energy range to reduce a selection effect.
 Monitor of All-sky X-ray Image (MAXI) is one of the X-ray instruments
 which have a capability to observe such an extremely soft GRB
 and alert its location to the community promptly
 \citep{2009PASJ...61..999M}.  

 MAXI is an experimental payload on the exposed facility of 
 Japanese Experiment Module attached to the International Space Station
 (ISS). It started nominal observation on August 15, 2009, 
 and keep monitoring the X-ray sky since then.
 MAXI has two scientific instruments: the Gas Slit Camera 
 \citep[GSC;][]{2011PASJ...63S.623M,2011PASJ...63S.635S} 
 and the Solid-state Slit Camera 
 \citep[SSC;][]{2010PASJ...62.1371T,2011PASJ...63..397T}. 
 GSC has a larger effective area and field of view (FOV)
 than those of SSC. Therefore, GSC is suitable to detect GRBs.
  In this paper, we present the results based on the GSC data.
 The GRB observations based on the SSC data will be presented elsewhere.

 MAXI employs a slit and slat collimator optics. 
 This optics has an advantage over conventional coded-mask systems
 in reducing the contamination from the cosmic X-ray background to 
 a point source.
 Therefore MAXI GSC achieves the highest sensitivity
 as a monitoring instrument in X-ray energy range, so far.

 One of the most comprehensive studies of GRBs below 10 keV has
 been accomplished by \citet{2005ApJ...629..311S}.
 They utilized the data sets observed by the Wide-field X-ray Monitor 
 \citep[WXM;][]{2003PASJ...55.1033S}
 on the High Energy Transient Explorer 2 
 \citep[HETE-2;][]{2003AIPC..662....3R}.  
 We will present here a comprehensive study of the MAXI/GSC GRB
 samples so that we can compare them with those of WXM.

\section{Observations and Data Analyses}

 \subsection{Gas Slit Camera(GSC)}

 The GSC consists of 12 
 one-dimensional position sensitive proportional counters operating in 
 the 2--30 keV range.
 Because the slit and slat collimator
 optics has a smaller FOV than that of a coded-mask
 optics, the cosmic X-ray background of GSC is rather lower than 
 HETE-2/WXM.
 Two GSC camera units, GSC-H (horizontal camera) and GSC-Z (zenithal
 camera),
 have an instantaneous FOV of 
 3$^{\circ}\times$160$^{\circ}$
 by pointing orthogonal directions each other.
 GSC covers 70\% of the whole sky every orbit.

\begin{figure}[htbp]
  \begin{center}
  \rotatebox{0}{
    \FigureFile(80mm,60mm){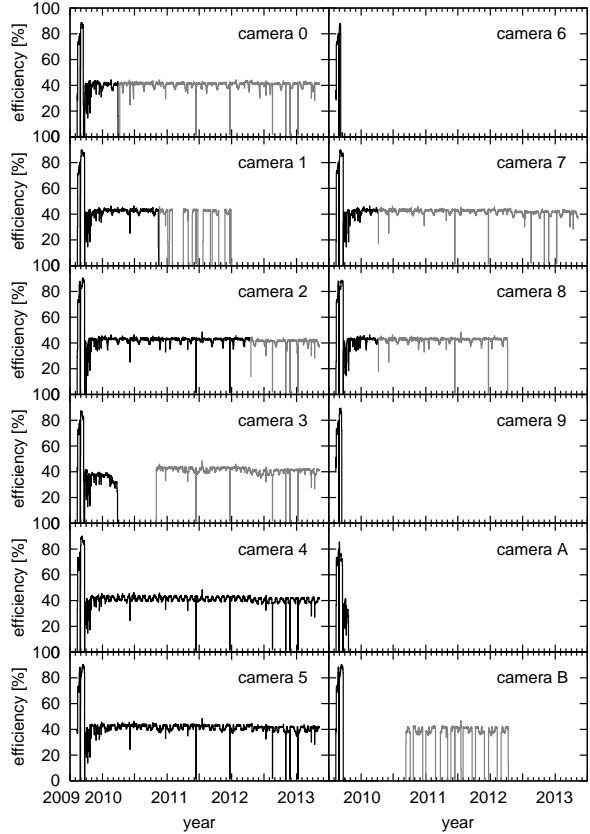}
   }
  \end{center}
  \caption{The operation efficiency of GSC counters.
     The black and gray lines represent the operation with 
     1650 V and 1550 V, respectively.
     Two cameras on the same row (e.g.,
     camera 0 and camera 6) cover the same FOV. Loss of either one of the
     cameras reduces the effective area to a half.
     }\label{fig:live}
\end{figure}

 The GSC counters are only operated within latitude of $\pm$ 40$^{\circ}$
 to avoid a risk of discharge due to a high particle background
 that could leave damages on the carbon-anode
 wires. Because of this restriction,
 the operation efficiency (i.e. fraction of actual observing time) of 
 GSC is about 40\% except for the first 1.5 months 
 (figure \ref{fig:live}).
 The counters which experience a discharge were tentatively stopped
 or operated by reducing voltage from 1650 V to 1550 V 
 (\citet{2011PASJ...63S.635S} in detail).
 The slight decrease of the 
 sensitivity is expected because of the reduction of the voltage.
 However, we find no significant difference in the observed rate of GRBs
 between the periods when the cameras are operated in 1650 V and 1550 V.

%

 MAXI scans a certain celestial position every 92 min that is the 
 orbital period of ISS.
 A typical transit time on a point source is about 40--100 sec,
 depending on the source-acquisition angle 
 $\beta$ \citep{2011PASJ...63S.635S}. 
 A typical effective area for 
 a point source is about 10 cm$^2$ in the 4--10 keV band.
 In addition, the effective area to a source varies during a scan,
 because the position of the source moves in the detector plane. 
 This variation 
 of the effective area during a scan makes it difficult 
 to know intrinsic variation of a source flux.
 However, we can estimate the uncertainty of the average flux
 during a scan by the method described in section \ref{sss:lc}.

 \subsection{Data reduction and sample selection}

  We analyzed X-ray event data of the GSC data process version 1.4.
  Both the data collected in 1650 V and 1550 V were used in the analysis. 
  For analyzing GRBs, we extracted X-ray events 
  within 10$^{\circ}$ from the best burst position.
  All the light curves and the spectra were 
  created from these extracted X-ray events.


 There are three ways  
 to identify GRBs or other transient events from the MAXI data. 
 First, the transient event can be identified automatically by the 
 search program
 called \textit{the MAXI Nova-Alert System} 
 \citep{2010fym..confP..63S}.  
 Second, the ground search of the MAXI data is possible by knowing the
 trigger time and the location of GRBs informed by other satellites.
 Third, we occasionally find transient sources 
 by eye inspection of daily or orbital all-sky images of MAXI.

 We selected transient events which had the signal-to-noise ratio (S/N)
 larger than 5 and only lasted for one scan. If the position of the event 
 matches within 1 degree to a known X-ray source,
 we exclude it from our sample.
 We also excluded the events which had galactic latitude $b$
 between $\pm 10^{\circ}$ to avoid contamination from the galactic 
 transients.
 The only exception is GRB 091230, which has a low galactic latitude but
 is confirmed as a GRB by INTEGRAL \citep{2009GCN..10298...1G}.

 In table \ref{tab:1}, we describe the parameters for
 35
 GRBs and short X-ray 
 transients observed by MAXI from August 2009 up to April 2013.
 The explanations of columns of table \ref{tab:1} are following.
 `Time' is the center time of the transit in which the burst was
 observed. The transit time depends on the source position.
 If the positions of the sources are determined by X-ray or optical
 telescopes accurately (table \ref{tab:ag}), 
 we used those position. Otherwise we used
 the position calculated with MAXI data (section \ref{ss:loc}).
 `(RA, Dec)' is the GRB location calculated with MAXI data in J2000,
 `loc. error' is error of the position, `cameras' is the camera numbers
 which observed the GRB, `S/N' is the signal-to-noise ratio of the GRB
 in the 2--20 keV band,
 `trigger' is the method that we found the GRB,
 `other sat.' is other satellites which observed the GRB.
 The trigger (or detection) time of the instrument from the time in 
 the `time' column are shown in the square brackets.
 `delay' is the time delay in sending GCN or ATel after the detection
 of the bursts.

 For the GRBs with X-ray and/or optical afterglows, we summarized
 information of the counterparts in table \ref{tab:ag}.
 `Band' column shows the band of observation of the counterparts.
 `(RA, Dec)' and error are the GRB location in J2000 and its error 
 determined by the observations of the counterparts.
 `GCN\#' is the number of GCN circular which report the position of
 X-ray or optical counterparts.
 `Redshift' is measured redshift (and its reference) of the GRB.

 \subsection{Localization}
 \label{ss:loc}

 A point spread function (PSF) of a constant source in the MAXI data can 
 be represented by
 the product of a spatial distribution in the detector anode direction
 and the effective area variation in the time direction.
 For an instantaneous observation, a spatial distribution of the 
 X-ray photons from a point source can be expressed as a Gaussian 
 distribution in the MAXI data.  
 The effective area to a source is expressed as a triangular function
 of the time.
 Therefore
 a PSF of a source can be represented by
 the product of the Gaussian in the detector anode direction
 and the triangular function in the time direction.

\begin{figure}[tbp]
  \begin{center}
   \rotatebox{0}{
    \FigureFile(75mm,80mm){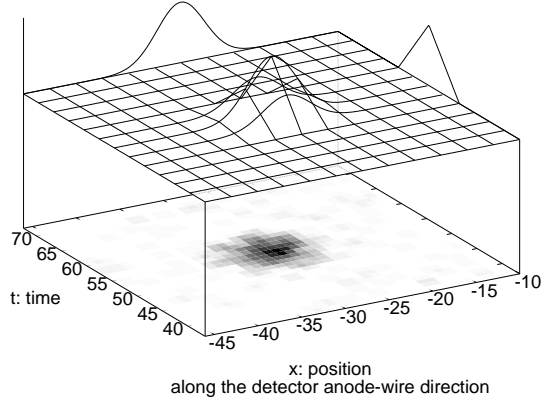}
    }
  \end{center}
  \caption{A schematic view of PSF fit of an X-ray source.
  The bottom map is the observed X-ray image of an X-ray source.
  We fit the PSF (the above function) to the image.
  An intersection of the PSF for the direction along the detector
  anode-wire ($x$) is modeled with a Gaussian. 
  The triangular shape of a intersection of the PSF for the direction 
  along the scan (i.e. time, $t$) reflects the variation of 
  the effective area during the scan. 
  The shape of intersections are also shown in the figure.
  The free parameters in fitting are position of the peak in the
  $x$--$t$ plane, which corresponds to the position in the celestial 
  coordinate, and the normalization, which corresponds to the
  source flux.
  }\label{fig:psf}
\end{figure}

 Figure \ref{fig:psf} shows a sample PSF of GSC.
 An intersection of the PSF for the direction along the detector
 anode-wire ($x$) is modeled with a Gaussian. 
 An intersection of the PSF for the direction along the scan 
 (i.e. time, $t$) reflects the variation of the effective area during 
 the scan, and thus has the triangular shape.

 In order to determine the position of an X-ray source,
 first we fit the two-dimensional (2-D) PSF 
 to the position histogram of X-ray events 
 in the $x$--$t$ plane. Then the position in $x$--$t$ plane is
 converted to the celestial coordinate using the information
 of the ISS attitude.

 During the position fitting process, we fix the width of 
 the Gaussian and the triangles 
 based on the source position at the detector plane. 
 There are three free parameters; the time at the peak, position at 
 the detector and normalization.

\begin{figure}[tbp]
  \begin{center}
   \rotatebox{0}{
    \FigureFile(75mm,80mm){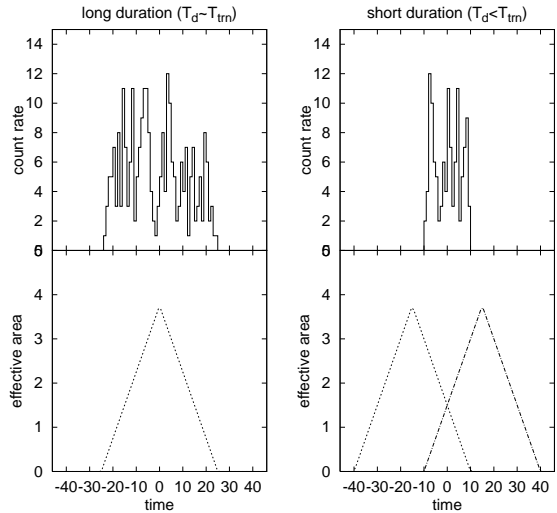}
    }
  \end{center}
  \caption{
  Illustrations of localization ambiguity for a
  long (left) and short (right) duration events. The solid
  lines are the simulated observed count rate curves and the dashed 
  lines or the dash-dotted lines are the effective area curves.
  When the event duration is short, the source position (time)
  cannot be determined as unique.
  }\label{fig:ea}
\end{figure}

 However, this PSF model is valid only for a source with constant flux.
 If the observed duration of the emission is significantly shorter than 
 the transit duration (i.e. width of the triangular function), 
 the position in the time direction becomes ambiguous. 
 We illustrate the situation schematically in Figure \ref{fig:ea}.
 For the case of long duration event (left panels), the peak time of the 
 triangular response can be determined without ambiguity.
 On the other hand, for the short event (right panels), the peak time
 can not be determined as unique.
 We illustrated possible two extreme cases in the panel: the 
 burst is observed at the end (dashed line) or beginning (dash-dotted 
 line) of the triangular response. The true response is somewhere
 between the two cases. 
 This ambiguity in the peak time reflects the ambiguity in
 the position of the source.

 In such a case, we calculate an error box of the source position 
 based on the ambiguity in time. At first, we calculate the position
 assuming a constant source
 (i.e. fit with the PSF for a constant source). We get a systematically
 small error box in this approach.

 Next, we extend the error box taking into account the ambiguity in the
 scan (time) direction.
 The size of the the error in this direction is 
 $\delta\theta (T_{\mathrm{trn}} - T_{\mathrm d})/T_{\mathrm{trn}}$,
 where $\delta\theta$, $T_{\mathrm{trn}}$, and $T_{\mathrm d}$ are
 the PSF size in the scan direction,
 the duration of the transit, and the observed duration of the burst,
 respectively.

 There are additional 0.1 deg systematic errors in the position 
 determination
 \citep{2011PhyE...43..692M}%
 \footnote{
 When the paper was written, the systematic error was 0.2 deg. 
 After the additional position calibration, 
 the systematic error has been reduced to 0.1 deg.
 }.

\begin{figure}[tbp]
  \begin{center}
   \rotatebox{0}{
    \FigureFile(75mm,80mm){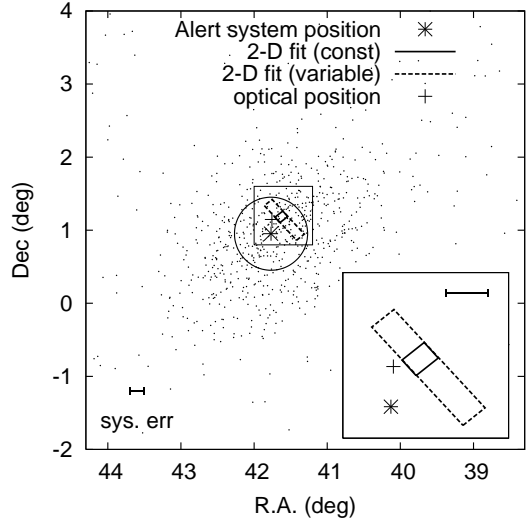}
    }
  \end{center}
  \caption{The localization error boxes of GRB 110213B.
  The dots are position of the X-ray events in the celestial coordinates.
  The star mark and the circle show the position and the error circle
  derived by \textit{the MAXI Nova-Alert System}, respectively . 
  The solid and dashed 
  boxes represent the positions derived by two-dimensional (2-D) source 
  fitting,
  where constant and variable flux of the source are assumed,
  respectively. The horizontal bars indicates the magnitude of the 
  systematic error for these error boxes. The position of the optical
  counterpart of this GRB is shown with the plus mark. 
  The square region is expanded in the inset.
  }\label{fig:1}
\end{figure}

  Figure \ref{fig:1} shows an example of localization error circles 
  and boxes of GRB 110213B. The small solid error box is the result
  of a fit under the assumption of PSF of a constant source.
  If we consider the shortness of the detection time, the error
  box is extended toward the direction of the scan, resulting
  in the dashed one.

\begin{figure*}[tbp]
  \begin{center}
   \rotatebox{0}{
    \FigureFile(160mm,80mm){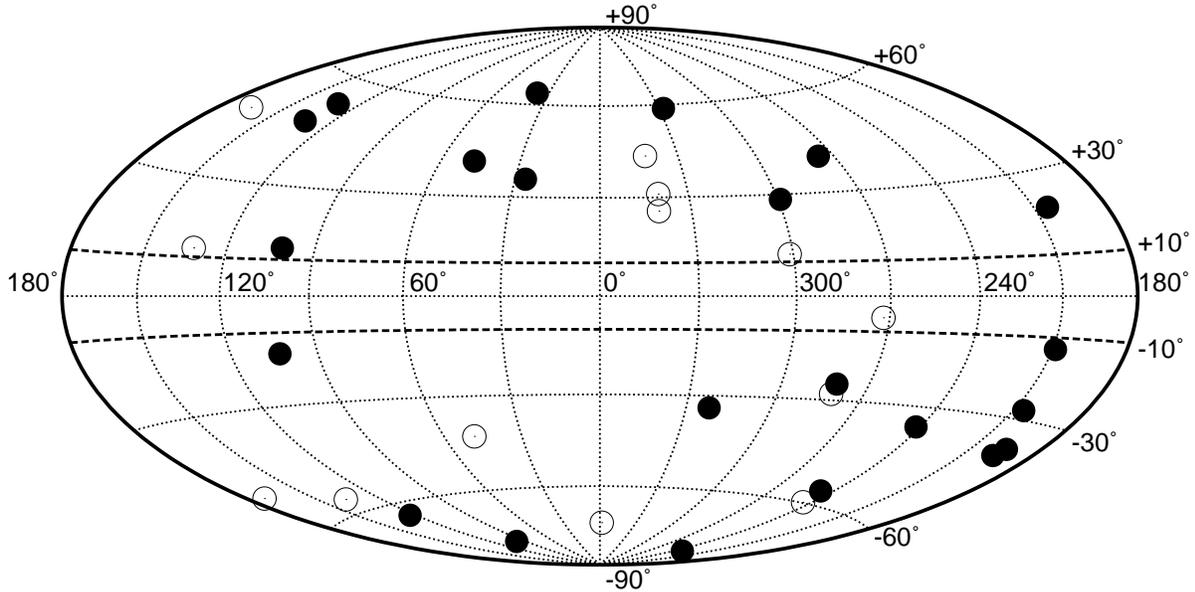}
    }
  \end{center}
  \caption{An all-sky map of MAXI GRBs in the galactic coordinates. 
   The open circles are the position of the GRBs also observed by
   other satellites. The filled circles are the position of the GRBs 
   observed only by MAXI. The thick dashed lines are at the galactic 
   latitude $b = \pm 10^{\circ}$, where we excluded the events
   (see text for the details).
  }\label{fig:allsky}
\end{figure*}

  The position determined only by the MAXI data is all consistent with
  the optical counterpart position by taking into account the systematic
  error.
  Figure \ref{fig:allsky} shows the all-sky map of the MAXI GRBs.

 \subsection{Light curves of the bursts}
 \subsubsection{Light curves of MAXI bursts}
 \label{sss:lc}
 To create light curves of MAXI bursts, we have to select the source 
 region in the image.
 In order to reduce systematic background variation, we adopted the data 
 selection using detector coordinates, rather than celestial coordinates.
 We selected the region that the separation of the source-acquisition 
 angle,
 which is the angle with respect to the orbital plane of MAXI
 \citep{2011PASJ...63S.623M},
 from the position of the source is $|\delta \beta| \leq 1.5^{\circ}$.
 The selected region projected to the celestial coordinate 
 forms a belt-like shape region with the width of 3$^{\circ}$.

 The light curves of MAXI data are as a result of a
 convolution of the source variation and a change of an effective area.
 In order to see the intrinsic source variation,
 we divide the observed counts by the effective area at the time.
 However, if a duration of an event is shorter than a transit time,
 a position of a source cannot be determined as mentioned in the previous
 section.
 In such a case, the effective 
 area and a flux variation of a source may also be ambiguous.

\begin{figure*}[tbp]
  \begin{center}
  \rotatebox{0}{
    \FigureFile(80mm,60mm){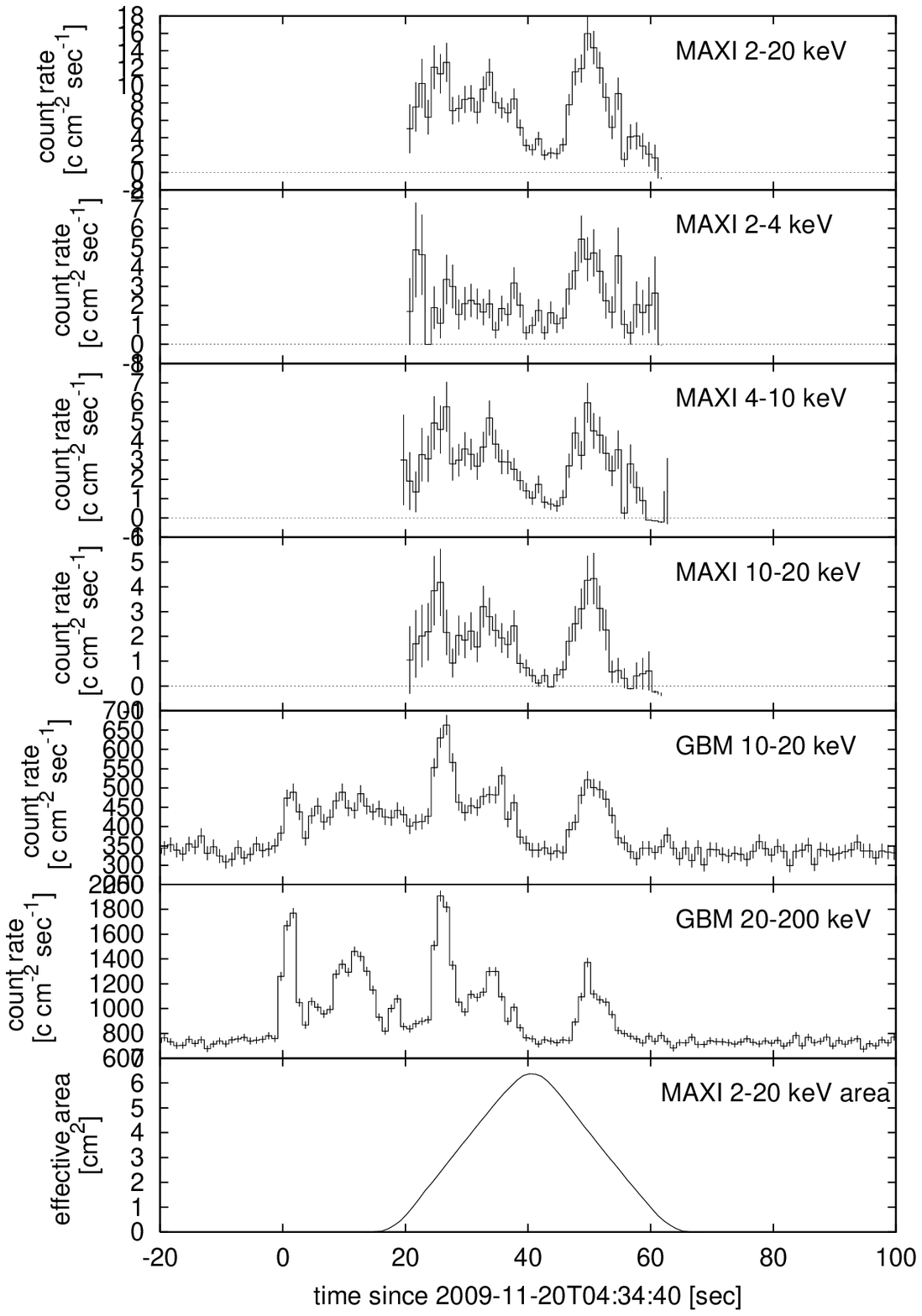}
    \FigureFile(80mm,60mm){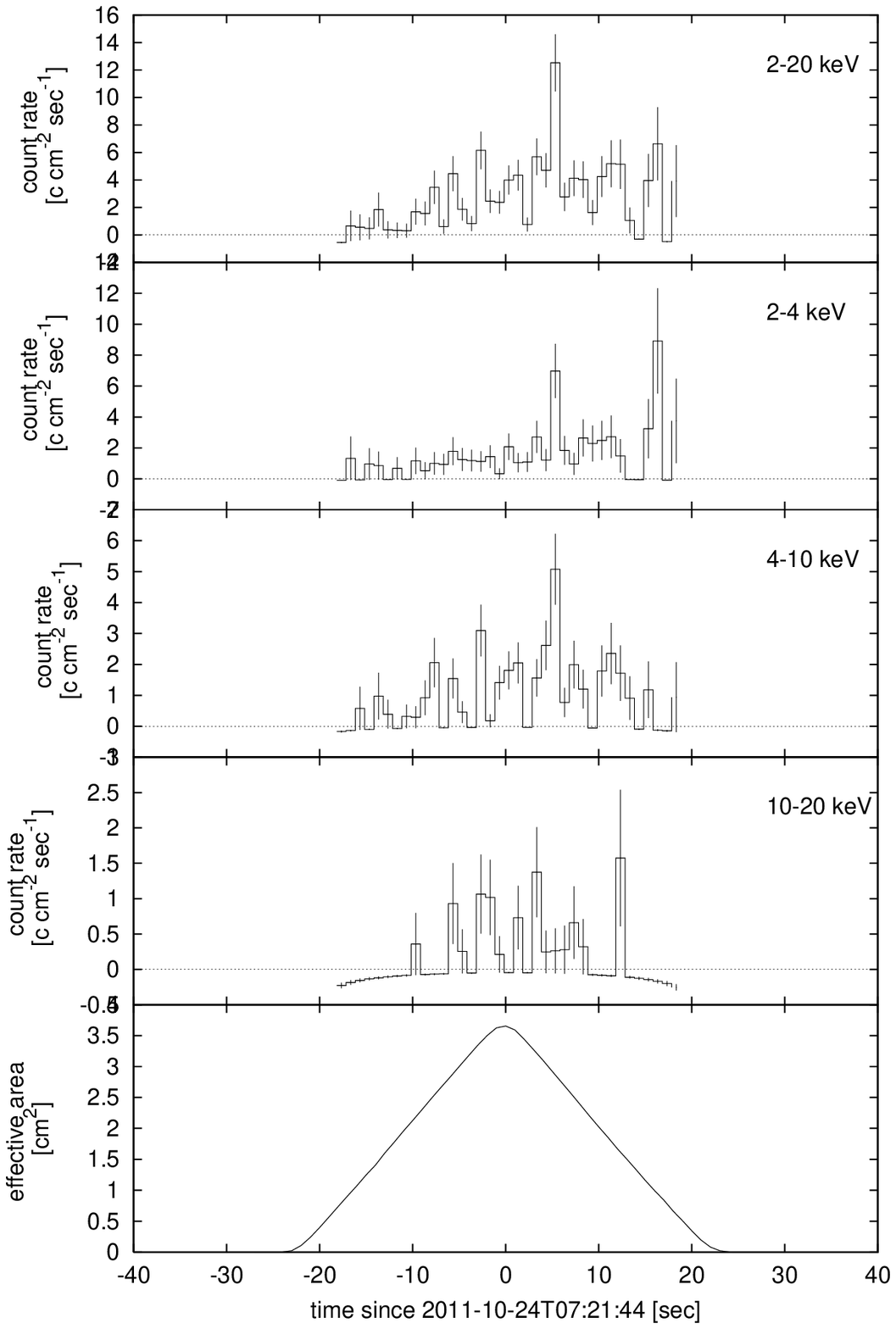}
   }
  \end{center}
  \caption{Light curves of GRB 091120 (left) and GRB 111024A (right). 
  The light curves of MAXI are
  corrected with the effective area of each energy band.
  For GRB 091120,
  the light curves observed by Fermi/GBM are also plotted for comparison.
  Effective area curves are shown in the bottom panels.
  }\label{fig:2}
\end{figure*}

 The left panels of figure \ref{fig:2} show samples of light curves of 
 GRB 091120 observed by MAXI, comparing with those of Fermi/GBM
 \citep{2009ApJ...702..791M}. 
 According to the light curves of Fermi/GBM,
 the duration of this burst is much longer than the transit time.
 Therefore the localization error is relatively small.
 We corrected the MAXI light curves with the effective area of each
 energy band. Note that we assumed the photon index of $-2$ when
 we calculate the effective area.
 The right panels of figure \ref{fig:2} shows the light curves
 of GRB 111024A, which is not observed by other satellites.
 The light curves are variable like classical GRBs.
 However, most of MAXI bursts are not so bright to observe the
 variability.
 The light curves of all MAXI bursts presented in this paper
 are available on the MAXI web page%
 \footnote{http://maxi.riken.jp/grbs}.

 \subsubsection{Hardness and intensity analyses}
 \label{ss:hr}
 Because the effective area of GSC can only achieve typically 
 $\sim 10$ cm$^2$,  GSC usually does not have enough photons to 
 perform spectral analysis for short-lived events.
  Therefore we decided to study
 two parameters, the average energy flux and the hardness ratio,
 which are possible to derive without performing spectral analysis with a
 relatively small assumption.  

 The hardness ratio is defined as the ratio of 
 photon flux in 8--20 keV band to that in 2--8 keV band,
 assuming a spectral slope $-2$%
 \footnote{
  We re-calculate the hardness using the spectral slope estimated from
  the hardness. However, the change due to the assumed spectral slope
  is smaller than the statistic error of the hardness
 }. 
 In order to calculate the average flux in the unit of 
 ergs cm$^{-2}$ s$^{-1}$, first we calculated the average photon flux
 in the 2--20 keV band.
 The average photon flux is determined as the total counts of the 
 bursts in the 2--20 keV band divided by the total effective area 
 times the scan duration around the burst. 
 The background count rate is calculated from the count rate
 before and after the transit.
 After the background subtraction,
 the total counts of the burst is calculated as the observed count
 during the transit.

 %
 To obtain reliable energy flux, we took into account the spectral 
 hardness of the burst in calculating the effective area. 
 We calculated a photon index in the simple power-law model
 from the observed hardness ratio. Then the photon flux is converted 
 into the energy flux using this estimated photon index.
 %

 %
 The uncertainty in the flux measurement comes not only from statistical 
 error but also from systematic uncertainty of the effective area.
 The degree of this systematic uncertainty depends on the ratio 
 of the observed duration of the burst $T_{\mathrm d}$ to the duration 
 of the transit $T_{\mathrm{trn}}$. 
 Since MAXI may not observe a whole GRB,
 $T_{\mathrm d}$ does not mean usual ``duration'' (like $T_{90}$), but
 it means the lower limit of the duration.
 In the case of $T_{\mathrm d} / T_{\mathrm{trn}} \geq 0.5$, the maximum
 uncertainty is 1.5 (i.e. the flux is underestimated and the true flux
 can be 1.5 higher than that of calculated from MAXI data). If 
 $T_{\mathrm d} / T_\mathrm{trn} < 0.5$,
 the maximum uncertainty in flux can be expressed as 
 \[
   \frac{T_{\mathrm{trn}}}{T_{\mathrm d}} - \frac{1}{2} \, ,
 \]
 which becomes larger for the shorter burst.

 In addition, there is possibility that the average flux here is
 different from the average flux observed by other satellites.
 This is because MAXI may observe only a part of long GRB.
 In fact, GRB 120711 became 5 times brighter than the flux at the time
 of MAXI observation after the end of MAXI transit.
 So, the average flux here should be treated with care.

\begin{figure}
 \begin{center}
  \FigureFile(80mm,80mm){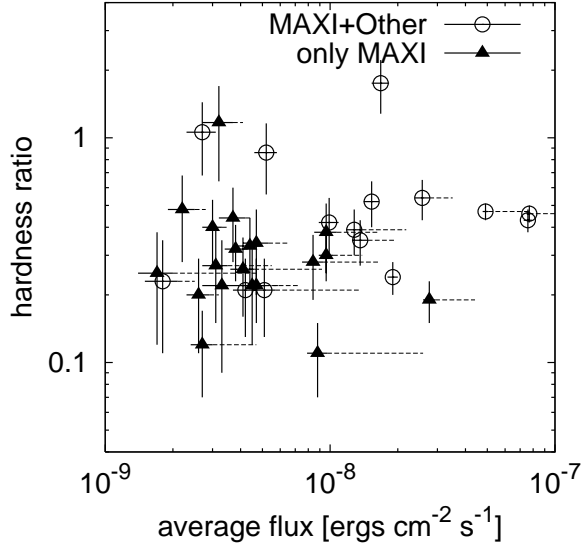}
 \end{center}
  \caption{Flux and hardness of MAXI GRBs. 
  The bursts which were also observed by other instruments
  are plotted with circles (simul.-GRBs).
  The bursts which were not observed by other instruments
  are plotted with triangles.
  The solid error bars correspond to the statistic errors and
  the dashed error bars on the average flux correspond to
  the systematic errors.
  }
  \label{fig:fh}
\end{figure}

 The flux and hardness parameters are summarized in table \ref{tab:2}.
 Figure \ref{fig:fh} shows the relationship between the hardness
 ratio and the time-averaged flux.
 The bursts only observed by MAXI (``only MAXI'', triangles) tend to
 distribute at the soft and the low flux region (lower left part of 
 the plot).
 On the other hand, most of the bright 
 ($> 10^{-8}$ ergs cm$^{-2}$ s$^{-1}$) bursts, which were also observed by
 other instruments simultaneously (``simul.-GRBs'', circles), 
 show relatively hard spectra.
 From this figure, we 
 conclude MAXI/GSC is generally sensitive to soft and dim bursts.
 Averages of the hardness of ``only MAXI'' and
 ``simul.-GRBs'' are 0.32 and 0.54, respectively.

 \subsection{Spectral analyses}
 \label{ss:spec}
  We selected bright GRBs with
  high S/N ($>$35) to have an enough statistics to perform spectral
  analysis.
  There are 7 GRBs which satisfy this criterion.
  If the bursts are simultaneously observed by Swift/BAT 
  \citep{2005SSRv..120..143B} 
  or Fermi/GBM
  \citep{2009ApJ...702..791M}, 
  we performed joint spectral fit of the MAXI spectrum and the spectrum 
  of those instruments.
  The spectrum of those high energy instruments is useful to constrain
  the broad-band spectral shape of GRBs.
  Since the MAXI transit time is rather short, the entire GRB episode
  can extend beyond the transit time.
  In these cases, we truncated the data of other missions into the same
  start and end time as MAXI's transit time for the spectral analyses.

  We analyzed the time averaged spectra of 7 GRBs. Four of them
  are jointly fitted with the GBM or the BAT data. We tested three type
  of models:
  power-law (PL), power-law with exponential cutoff (CPL), and GRB model
  (GRBM) \citep[so-called Band Function;][]{1993ApJ...413..281B}.
  When we perform the spectral analyses without the GBM or the BAT data,
  the $E_{\mathrm{peak}}$ may converge at an inappropriate value due to the
  limitation of the GSC energy range. 
  In such a case, we calculated lower limit of $E_{\mathrm{peak}}$ 
  at 90\% confidence by fixing a low energy photon index $\alpha$ to
  $-1.0$.
  When a high energy photon index
  $\beta$ is not well constrained, 
  we fix the index $\beta$ to $-2.3$.
  We do not consider these models with artificially fixed parameters
  as the best fit models.
  The results are summarized in table \ref{tab:spec}.  
  All the errors in the table are in 90\% confidence.  
  Although the S/N of GRB 090926B is lower than our criteria,
  we also listed the spectral parameters of GRB 090926B in the table, 
  because they are already given by \citet{2011PASJ...63S1035S}.
  The $E_{\mathrm{peak}}$ of the MAXI-Fermi GRBs are 60--100 keV
  range. While the $E_{\mathrm{peak}}$ of the MAXI-Swift and 
  MAXI GRBs may be located below 20 keV.

\section{Discussions}

 \subsection{GRB rate and detection sensitivity}
 
 Conventional studies of GRB rate are targeted on relatively high energy 
 band.
 For example, \citet{2002ApJ...578..304S} made a log $N$ -- log $P$ curve
 in the 50--300 keV band using BATSE and \textit{Ulysses} samples.
 \citet{2009aaxo.conf..224S} estimated that
 the detection limit of MAXI/GSC is equivalent to 0.4 
 photons cm$^{-2}$ s$^{-1}$ in the 50--300 keV band
 based on the simulation study.
 The corresponding rate is about 400 GRBs per year.
 An average 
 observing efficiency of MAXI/GSC is about 40\% and sky coverage
 is about 2\% of the whole sky.
 Multiplying these numbers, we expect $\sim$ 3 events per year for the 
 MAXI GRB rate. However we have observed 35 event in 
 44 months, which is more than three times higher than this estimation.
 Interestingly, the rate of ``simul.-GRB'' 
 is close to the expected rate. 
 Since the two thirds of MAXI GRBs are probably not observed by other 
 satellites,
 those MAXI GRBs
 are not included in the calculation of the expected number
 by \citet{2009aaxo.conf..224S}.
 According to figure \ref{fig:fh}, most of GRBs in hardness $>$ 0.4 are
 ``simul.-GRB''.
 On the other hand, ``only MAXI'' bursts dominate the range of
 hardness $<$ 0.4 and also tend to be underluminous. 
 Therefore, it is reasonable that GRB rate expected from the 
 log $N$ -- log $P$ based on BATSE and Ulysses samples does not agree 
 with the observed rate in the MAXI/GSC energy band.
 This fact suggests that there are many soft bursts 
 failed to be detected by the
 traditional GRB instruments.

\begin{figure}
 \begin{center}
  \FigureFile(80mm,80mm){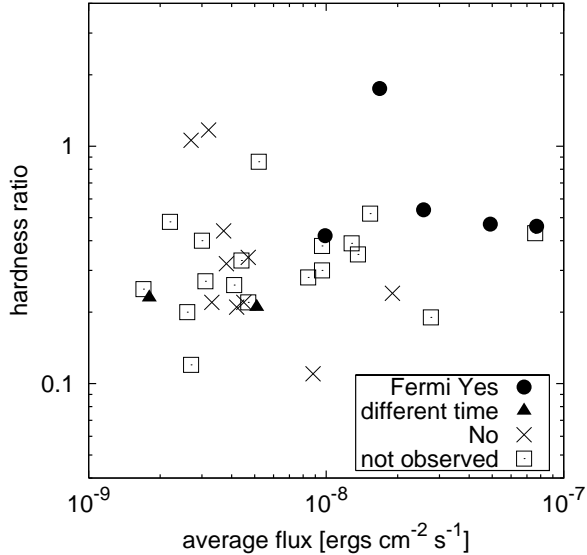}
 \end{center}
  \caption{Flux and hardness of MAXI GRBs with or without Fermi/GBM
  detection. 
  The bursts which are also observed by GBM
  are plotted with circles. Two events shown by triangles
  were detected by GBM, but the trigger time were not
  during the MAXI transit (see text).
  The bursts shown with cross marks were observed but no
  significant signal is found.
  The bursts shown with squares were not observed (occulted
  or occurred during the off time) by GBM.
  }
  \label{fig:fermi}
\end{figure}

 There is a convincing evidence of existence of soft and underluminous 
 GRBs.
 According to \citet{2009ApJ...702..791M}, Fermi/GBM operates 90 \% of
 the time. About 50 \% of MAXI events are occulted by the Earth
 for Fermi/GBM.
 We confirmed that 17 out of 35 MAXI GRBs were inside the Fermi/GBM FOV 
 and not occulted by the Earth.
 However, only 7 GRBs are detected by Fermi/GBM.
 Thus, for the other 10 events,
 we analyzed the daily monitoring CSPEC data, which are publicly
 available in Fermi Science Support Center web page%
 \footnote{http://fermi.gsfc.nasa.gov/ssc/}.
 We can not find any significant signal around MAXI trigger time
 for all those 10 events.
 Figure \ref{fig:fermi} shows the flux and hardness of the GRBs
 with or without Fermi/GBM detection.
 GRB 110426A and GRB 120908A, plotted with triangles, 
 were also triggered by GBM, but
 the trigger time were before the MAXI observations 
 (cf. table \ref{tab:1}).
 At the time of MAXI observation, the signal in the GBM data 
 is not apparent.
 There is clear trend in the figure. Fermi/GBM detected
 bright and hard events among the MAXI GRBs.

 GRB 101117A was in the FOV of Swift/BAT, but no significant signal 
 was seen in the BAT data.
 For Suzaku/WAM \citep{2009PASJ...61S..35Y}, 
 only GRB 090831 has been detected. Out of the other
 34 GRBs, 19 were not occulted by the Earth nor occurred during 
 the off time.
 Based on those studies, we concluded that large fraction of
 MAXI GRBs are not detectable by the traditional GRB instruments.

\begin{figure*}[t]
 \begin{center}
  \FigureFile(80mm,80mm){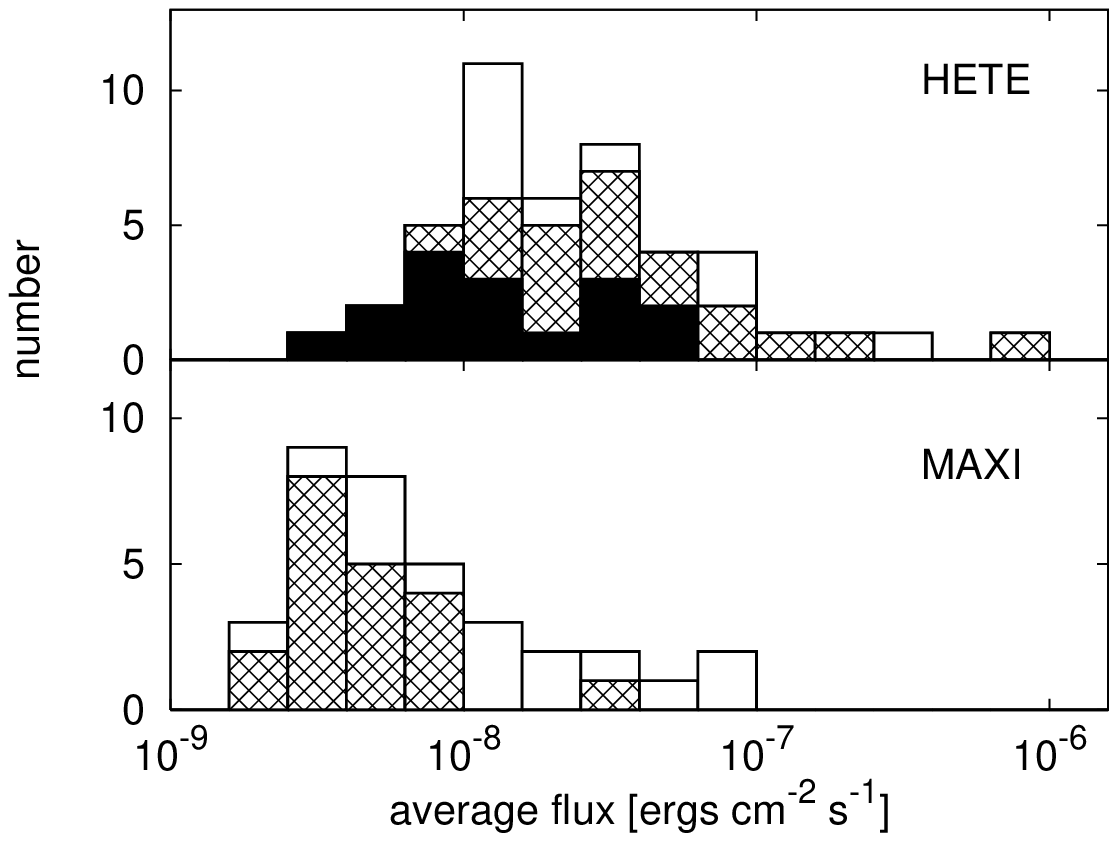}
  \FigureFile(80mm,80mm){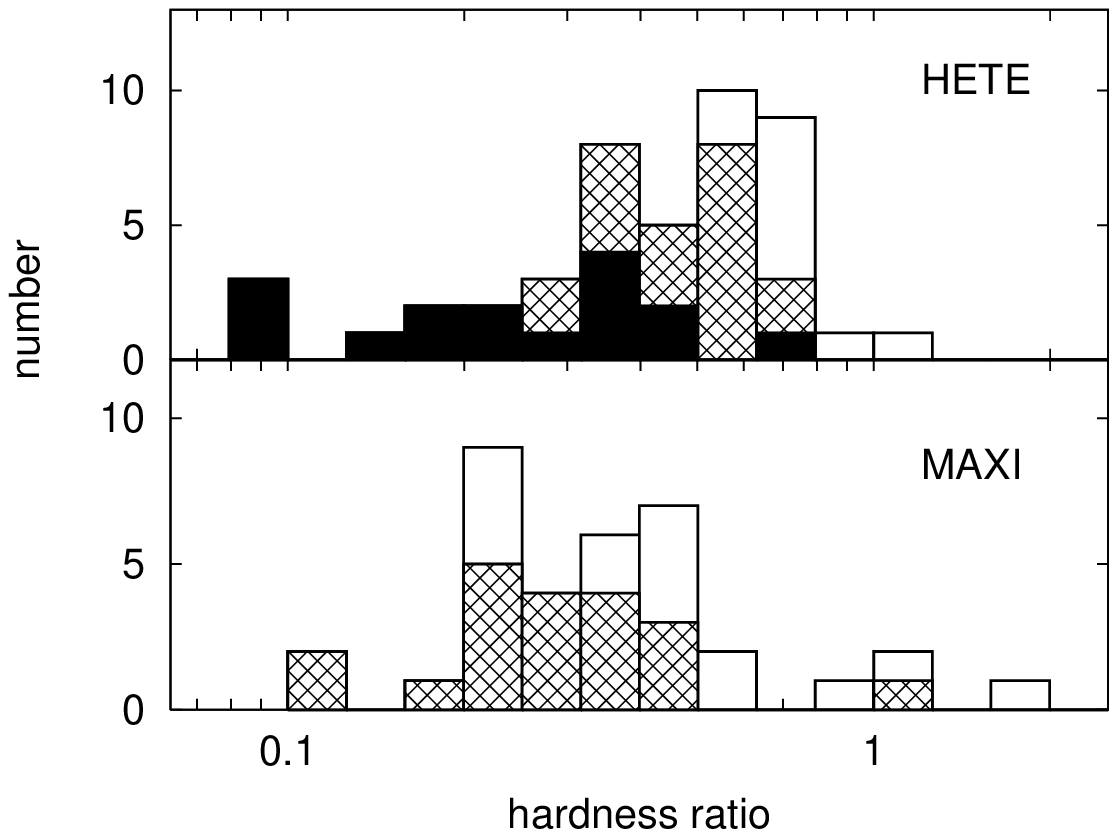}
 \end{center}
  \caption{Histograms of time-averaged flux in the 2-20 keV band (left) 
  and hardness (right) 
  distribution of MAXI and HETE-2 GRBs. 
  The hardness is defined as the ratio of 
  photon flux in the 8--20 keV band to the 2--8 keV band. 
  The histograms of top panels are result
  of HETE-2/WXM \citep{2005ApJ...629..311S}.
  The histograms of MAXI/GSC are plotted in the bottom panels.
  The GRBs observed by HETE-2 are classified into (classical) GRBs (open), 
  X-ray rich GRBs (hatched), and X-ray flashes (filled).
  The hatched bursts in the bottom panels are ``only MAXI''
  bursts.
  }
  \label{fig:hist}
\end{figure*}

 In order to compare the results with observations in similar energy
 range, we plotted histograms of flux and hardness distribution of
 GRBs observed by HETE-2/WXM \citep{2005ApJ...629..311S} in figure 
 \ref{fig:hist}.
 From the left panel, we see 
 that the average flux of the MAXI bursts are systematically lower 
 than that of the WXM GRBs. 
 Although the effective area of GSC is 3--6 times smaller than that of 
 WXM, the slit and collimator optics of GSC makes the background lower 
 and thus achieves higher sensitivity than WXM.
 The right panels of figure \ref{fig:hist} show the distribution
 of the hardness of the MAXI and the HETE samples. 
 The hardness distribution of the MAXI GRBs,
 especially ``only MAXI'' bursts, has a similar trend to that of
 the HETE XRFs. We can compare the mean hardness value of each
 GRB class.
 the HETE XRFs, X-ray rich GRBs, and classical GRBs have
 the mean hardness of 0.29, 0.49, and 0.75, respectively.
 The mean hardness of ``only MAXI'' bursts is 0.32, and it is
 the nearest to the HETE XRFs.

 \subsection{Spectral properties of MAXI bursts}

  MAXI has unique capability to observe soft GRBs.
  In HETE-2 samples,  about one third of GRBs is classified into XRFs
  \citep{2005ApJ...629..311S}.  Although it is difficult to classify
  GRBs solely from the MAXI data, we expect that roughly one third of 
  the MAXI GRBs are classified into XRFs on the basis of the
  similarity between the distributions of the hardness of MAXI and HETE
  bursts.

\begin{figure}
 \begin{center}
  \FigureFile(80mm,80mm){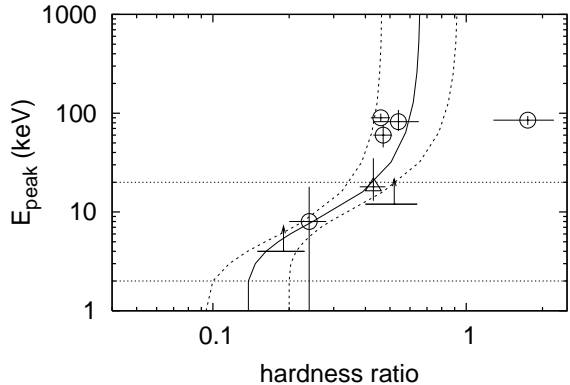}
 \end{center}
  \caption{The relation between hardness ratio (8--20 keV/2--8keV)
  and $E_{\mathrm{peak}}$ of MAXI GRBs.
  The solid line shows the correlation calculated with Band function
  with spectral indices $\alpha = -1.0$ and $\beta = -2.3$.
  The right and left dashed curves correspond to 
  a harder ($\alpha = -0.7$, $\beta = -2.0$) or 
  a softer ($\alpha = -1.3$, $\beta = -2.6$) spectral models,
  respectively.
  The open circles are results from
  the joint spectral fitting with other instruments.
  GRB 110213B is plotted with the triangle. 
  For GRB 111024A and GRB 120510A, we cannot obtain reasonable 
  $E_{\mathrm{peak}}$ without fixing $\alpha$.
  Therefore lower limits calculated with fixed $\alpha = -1$ 
  are shown with arrows.
  The horizontal dotted lines show the upper and
  lower boundaries of GSC energy range used for the hardness calculation.
  }
  \label{fig:ep}
\end{figure}

  A traditional parameter ,$E_{\mathrm{peak}}$,
  is used to represent the softness of GRBs.
  However, only a few bursts have enough statistics 
  for spectral analysis in the MAXI samples.  
  Instead of $E_{\mathrm{peak}}$
  we calculated hardness of the bursts
  (section \ref{ss:hr}).
  In order to examine the relationship between the $E_{\mathrm{peak}}$
  and  the hardness,
  we calculated the hardness ratio in the MAXI energy bands,
  photon flux in the 8--20 keV band to the 2--8 keV band,
  using the Band function with fixed 
  indices $\alpha = -1.0$ and $\beta = -2.3$.  The result is shown in
  figure \ref{fig:ep}.  
  According to \citet{2006ApJS..166..298K}, the distributions of
  indices $\alpha$ and $\beta$ have deviation of $\sim 0.3$.
  Considering this deviations, we also plotted the curves of
  a harder ($\alpha = -0.7$, $\beta = -2.0$) and
  a softer ($\alpha = -1.3$, $\beta = -2.6$) cases
  in dashed lines.
  In this figure, we also plotted the best-fit 
  $E_{\mathrm{peak}}$ or its lower limits of 8 MAXI GRBs shown in table \ref{tab:spec}.
  Although the uncertainty of 
  the $E_{\mathrm{peak}}$ obtained form the spectral analyses are large, 
  they are consistent with the $E_{\mathrm{peak}}$ inferred from the hardness.
  The exception is GRB 090926B locating right with hardness $>1$.
  Since the spectral index $\alpha$ of this burst is positive 
  \citep{2011PASJ...63S1035S}
  and far from the assumed value $-1$,
  it is not surprising that this GRB does not follow the relationship.

  In figure \ref{fig:hist},
  we see most of the ``only MAXI'' events have hardness $< 0.4$. 
  According to figure \ref{fig:ep}, the hardness of $< 0.4$
  corresponds to $E_{\mathrm{peak}} < 20$ keV. 
  Since all the bursts which have $E_{\mathrm{peak}} < 20$ are 
  classified to XRFs \citep{2005ApJ...629..311S,2008ApJ...679..570S},
  most of the ``only MAXI'' events can be classified as XRFs.

 \subsection{What does MAXI observe?}

  An essential question is whether the ``only MAXI'' events are XRFs
  or other transient phenomena.
  A reliable way to answer the
  question is to carry out follow-up observations of these bursts
  in other wavelengths.
  However we have not succeeded them yet. 
  We do not know
  redshift for any of the ``only MAXI'' events so far.

\begin{figure}
 \begin{center}
  \FigureFile(80mm,80mm){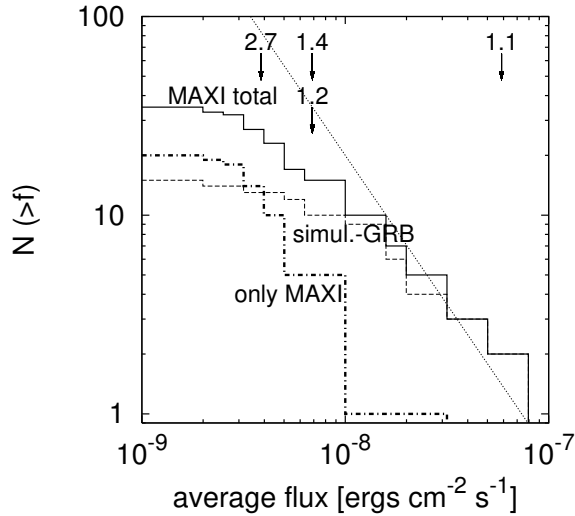}
 \end{center}
  \caption{The cumulative distribution of average flux.
  The solid, dashed, and dash-dotted lines represent the numbers of total,
  ``simul.-GRBs'', and ``only MAXI'' events, respectively.
  The dotted line has a slope of $-3/2$, 
  indicating a uniform distribution.
  The arrows show the fluxes of four GRBs with known redshifts. 
  The numbers on the arrows are their redshift values.
  }
  \label{fig:ncum}
\end{figure}

  Instead of direct study of the distribution of distance to the
  source, we plotted cumulative distribution of the average flux in figure
  \ref{fig:ncum}.  In the figure, we also indicated the fluxes of four
  bursts with known redshifts, which are ``simul.-GRBs''.
  The best fit slope to the observed distribution of ``simul.-GRBs''
  becomes flatter than $-3/2$ below the flux around $10^{-8}$
  ergs cm$^{-2}$ s$^{-1}$, while the slope of ``only MAXI'' bursts 
  is close to $-3/2$ down to the flux range of $3 \times 10^{-9}$
  ergs cm$^{-2}$ s$^{-1}$.
  It would suggest that ``only MAXI'' burst has a uniform distribution 
  down to $3 \times 10^{-9}$ ergs cm$^{-2}$ s$^{-1}$ 
  and there is no 
  selection effect of the triggering sensitivity
  above this level.
  Generally the flattening of the slope from $-3/2$ at the low 
  flux is interpreted as a cosmological effect
  \citep{1992Natur.355..143M}. 
  Therefore the difference in the slope of ``only MAXI'' GRBs suggests 
  the intrinsically different population from ``simul.-GRBs''.
  The ``only MAXI'' GRBs have low luminosity and distribute
  closer to us than the ``simul.-GRBs''.
  \citet{2004PhDT.........1M} inferred that XRFs observed by HETE-2
  distribute closer distance than hard GRBs, because the slope of the
  HETE-2 log $N$ -- 
  log $P$ distribution of XRFs are steeper than that of hard GRBs
  above 2 photons cm$^{-2}$ s$^{-1}$ .
  Our MAXI result shows similar trend to the results of HETE-2.

\begin{figure}
 \begin{center}
  \FigureFile(80mm,80mm){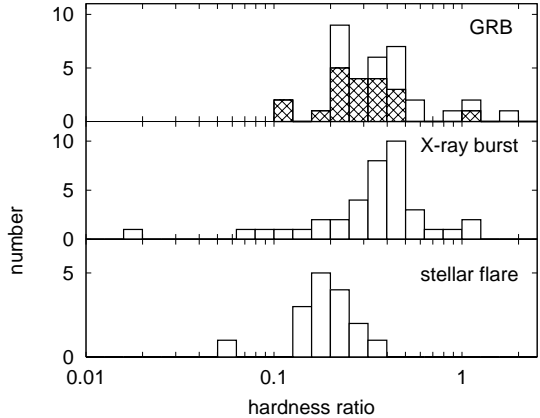}
 \end{center}
  \caption{The distribution of the hardness of GRBs, XRBs, and 
   stellar flares observed with MAXI.
  The hardness is defined as the ratio of 
  photon flux in the 8--20 keV band to the 2--8 keV band. 
  The hatched events in the top panel indicate ``only MAXI''
  bursts.
  }
  \label{fig:hrgal}
\end{figure}

  Is there any possibility that 
  ``only MAXI'' events are galactic transients?
  In order to investigate whether we can distinguish galactic transients
  from GRBs by hardness ratio or not, we calculated the hardness of
  galactic transients assuming their typical emission models.
  As sources of galactic transients, we considered (type-I) X-ray bursts
  (XRBs), which are thermonuclear flashes on neutron stars,
  and stellar flares. 
  We plotted the distribution of hardness of GRBs and galactic transients
  in figure \ref{fig:hrgal}.
  The hardness range of GRBs is overlapping with those of XRBs or stellar 
  flares.
  Therefore it is difficult to 
  distinguish between GRBs and XRBs only from the hardness ratio.
  The most reliable way to distinguish GRBs from XRBs or stellar flares
  is to refer the catalogs of known sources.
  However this method is not applicable to a new transient source.
  An example of such event is an X-ray burst from Swift~J1741.5-6548
  \citep{2013ATel.4911....1N}. MAXI detected a burst on 2012 December
  25, and the position of the burst did not match with any known 
  X-ray source. We reported the event as GRB 121225A at first
  \citep{2012GCN..14100...1O}. Two months later, Swift detected a
  previously unknown transient source \citep{2013ATel.4902....1K}, and 
  the position of the source was marginally consistent with
  the position of GRB 121225A.  If a follow-up observation of 
  GRB 121225A had been performed immediately after the MAXI detection, 
  we would have found the counterpart.
  We learned the importance of immediate follow-up observations from this example.

  There are possibilities of confusing XRBs from previously unknown source
  like the example of Swift~J1741.5-6548 in the MAXI GRB samples.
  In fact, HETE~J1900.1-2455 was not known when HETE-2 detected the
  first XRB from this object
  \citep{2005ATel..516....1V,2007PASJ...59..263S}.
  Although a
  persistent emission from the
  XRB source is expected,
  MAXI is often not sensitive enough to detect a weak persistent 
  emission.
  Therefore it is desirable to carry out follow-up observations 
  by a highly sensitive X-ray telescope
  to find a persistent emission of a MAXI transient event 
  to distinguish GRBs from XRBs.

  Unlike GRBs or XRBs, stellar flares have a longer time scale than a 
  scan interval of MAXI. So they are usually observed for 
  multiple scans. 
  Therefore they are easy to be distinguished
  from GRBs or XRBs.
  Weak stellar flares may be observed only at the peak due to 
  the sensitivity limit of MAXI. 
  However, the hardness
  distribution of stellar flares (figure \ref{fig:hrgal}) peaks at 
  around the
  lowest end of the hardness distribution of ``only MAXI''.
  Therefore, we can distinguish the ``only MAXI'' events from stellar 
  flares by the temporal and the spectral information.
  We believe that unknown stellar flares are less likely to be confused
  in our MAXI GRB samples.

\section{Conclusion}

  We selected bright transient events in high ($> |10^{\circ}|$) 
  galactic latitude 
  from the MAXI/GSC data. We calculated
  the average energy flux and hardness of these events. The results
  show that the bursts observed only by MAXI (``only MAXI'') tend to 
  have soft spectra and relatively low flux, comparing to
  GRBs simultaneously observed by other satellites (``simul.-GRBs'').

  In comparison with the GRBs observed by HETE-2/WXM,
  the average flux of MAXI bursts are lower than those of HETE-2,
  while the effective area of GSC is 3--6 times smaller than that of WXM.
  This better sensitivity of GSC comes from its low background,
  which is a benefit of the slit and collimator optics.
  From the distributions of the hardness, we found 
  that ``only MAXI'' bursts have the similar distribution of 
  XRFs observed by HETE-2/WXM.

  Since most of MAXI bursts do not have enough statistics to 
  perform spectral analysis,
  we compared the measured hardness with the calculated 
  hardness assuming the standard GRB spectral parameters. 
  As a result, we found most of the ``only MAXI'' events have 
  hardness $< 0.4$, which corresponds to $E_{\mathrm{peak}} < 20$ keV. 


  As of now, there is no ``only MAXI'' burst with known redshift.
  The log $N$ -- log $S$ slope of ``only MAXI'' bursts is close to 
  $-3/2$,
  while ``simul.-GRBs''
  show the slope shallower than $-3/2$.
  The difference of the slope of ``only MAXI'' GRBs 
  suggests that they are intrinsically dim and distribute closer
  than ``simul.-GRBs''.

  We looked into a possibility that there was a confusion of galactic
  transients to the MAXI GRBs. Since the distribution of the hardness of 
  X-ray bursts and stellar flares
  overlap with those of MAXI GRBs, 
  it is difficult to classify these transients solely by the hardness.
  The above facts suggest that it is 
  essential to carry out follow-up observations and find counterparts.
  Direct measurement of redshift distribution and/or identifying a
  counterparts in other wavelengths are 
  needed to unveil those events.

\bigskip


This research was partially supported by the Ministry of Education, 
Culture, Sports, Science and Technology (MEXT), Grant-in-Aid No.19047001, 
20041008, 20244015, 21340043, 23740147, 24684015, 24740186, and 
Global-COE from MEXT ``Nanoscience and Quantum Physics''.


\bibliographystyle{aa}
\bibliography{mybib}

\clearpage

\begin{table*}[p]
\begin{center}
\caption{A summary of GRBs and short X-ray transient observed by MAXI}
\label{tab:1}
\scriptsize
\begin{tabular}{llD{.}{.}{3}D{.}{.}{3}rD{x}{\times}{3}cD{.}{.}{1}ccr}
\hline
\hline
  GRB name & \multicolumn{1}{c}{time\footnotemark[$*$]} & 
  \multicolumn{2}{c}{(RA, Dec)} & 
  \multicolumn{2}{c}{loc. error\footnotemark[$\dagger$]} 
  & cameras\footnotemark[$\ddagger$] & 
  \multicolumn{1}{c}{S/N}
  & trigger\footnotemark[$\S$] & other sat.\footnotemark[$\|$]
  &  delay\footnotemark[$\#$] \\
\hline

  090831                    & 2009-08-31 07:37:07 & 145.4  &   +51.4   &
   C & \timeform{60'}                 & 2,8   &  185.3  & O & 
                                     F[$-31$](1)/K[$-33$](2)/ & 12h \\
   &&&&&&&&&                                      W[$-31$](3) &     \\
  090926B                   & 2009-09-26 21:55:39 &  46.3  &   -39.1   &
   C & \timeform{60'}                 & 1,7   &   28.3  & O & 
                                       S[$+9$](4)/F[$-10$](5) &  9h \\
  091012\footnotemark[$**$] & 2009-10-12 10:25:51 & 182.82  &  +63.37  &
   C & \timeform{12'}                 & 0,1,7 &   14.6  & H & &     \\
  091120                    & 2009-11-20 04:35:20 & 226.81  &  -21.79  &
   C & \timeform{30'}                 & 1,7   &  272.1  & O & 
                                      F[$-40$](6)/K[$-44$](7) & 97h \\
  091201                    & 2009-12-01 21:48:36 & 118.6   &  +16.6   &
   C & \timeform{24'}                 & 3     &   19.0  & M & & 19h \\
  091230\footnotemark[$**$] & 2009-12-30 06:27:15 & 132.91  &  -53.88  &
   C & \timeform{21'}                 & 2,8   &    9.2  & O & 
                                                  I[$+15$](8) &     \\
  100315A                   & 2010-03-15 17:13:40 &  74.95  &   -6.63  &
   B & \timeform{166'}x\timeform{45'} & 2,8   &    8.0  & M & & 14h \\
  100327                    & 2010-03-27 17:08:20 & 346.03  &  +42.90  &
   B & \timeform{58'}x\timeform{28'}  & 1,7   &   12.4  & M & & 66h \\
  100415A                   & 2010-04-15 03:44:54 &   7.48  &  -15.57  &
   B & \timeform{104'}x\timeform{36'} & 4     &   26.6  & M & &  9h \\
  100510A                   & 2010-05-10 19:27:23 & 355.8   &  -35.6   &
   B & \timeform{83'}x\timeform{16'}  & 1     &   51.2  & M & 
                                                  F[$-16$](9) & 24h \\
  100616A                   & 2010-06-16 01:42:17 &  50.95  &  -40.62  &
   B & \timeform{110'}x\timeform{70'} & 4,5   &   14.0  & M & &  8h \\
  100701A                   & 2010-07-01 06:54:31 & 188.86  &  -34.26  &
   B & \timeform{122'}x\timeform{21'} & 5     &   32.5  & M & &  5h \\
  100823A                   & 2010-08-23 17:25:52 &  20.70  &   +5.84  &
   C & \timeform{7'}                  & 4,5   &   67.4  & O & 
                                                  S[$-17$](10)& 95h \\
  100911                    & 2010-09-11 14:58:24 & 103.41  &  -70.43  &
   B & \timeform{39'}x\timeform{15'}  & 1,2   &    8.7  & H & & 41h \\
  101117A                   & 2010-11-17 07:32:57 &  89.63  &   -2.30  &
   B & \timeform{38'}x\timeform{22'}  & 0     &   28.4  & M & &  3h \\
  101210\footnotemark[$**$] & 2010-12-10 03:38:27 &  61.66  &   -5.36  &
   C & \timeform{20'}                 & 4,5,B &    8.8  & M & &     \\
  110213B                   & 2011-02-13 14:32:08 &  41.76  &   +1.15  &
   B & \timeform{40'}x\phantom{1}\timeform{7'} 
                                      & 4     &  139.8  & M & 
                                                 K[$-35$](11) & 14h \\
  110402\footnotemark[$**$] & 2010-04-02 02:33:55 &  62.52 &    -3.00  &
   E & \timeform{30'}x\timeform{20'}  & 0,7   &   11.6  & M & &     \\
  110426A                   & 2011-04-26 15:08:35 & 221.18 &   -10.78  &
   B & \timeform{262'}x\timeform{16'} & 4,5   &   20.2  & M & 
                                                F[$-128$](12) &  9h \\
  110916                    & 2011-09-16 20:33:12 & 171.68 &   -17.77  &
   B & \timeform{77'}x\timeform{19'}  & 2,8   &   13.8  & M & & 85h \\
  111024A                   & 2011-10-24 07:21:44 & 221.93 &   +25.87  &
   B & \timeform{12'}x\phantom{1}\timeform{9'} 
                                      & 4     &   74.8  & M & &  5h \\
  120424A                   & 2012-04-24 16:47:29 &  23.985 &  -29.87  &
   C & \timeform{16'}                 & 4,5   &   17.1  & M & & 19h \\
  120510A                   & 2012-05-10 08:48:06 &  44.285 &  +72.850 &
   C & \timeform{10'}                 & 0,7   &   52.8  & M & 
                         K(13)\footnotemark[$\dagger\dagger$] &  5h \\
  120528B                   & 2012-05-28 18:12:08 &  77.59  &  -37.80  &
   B & \timeform{78'}x\timeform{22'}  & 2,7   &   26.5  & M & 
                                                 K[$-20$](14) &  8h \\
  120528C                   & 2012-05-28 21:20:45 &  12.93  &   -0.95  &
   E & \timeform{48'}x\timeform{36'}  & 4,5   &    6.9  & H & & 31h \\
  120614A                   & 2012-06-14 05:49:10 & 312.73  &  +65.16  &
   C & \timeform{10'}                 & 0,7   &   33.2  & M & &  2h \\
  120622A                   & 2012-06-22 03:21:51 & 205.43  &   -1.71  &
   B & \timeform{130'}x\timeform{30'} & 2     &   12.1  & M & &  2h \\
  120626B                   & 2012-06-26 13:38:12 & 175.77  &  +68.50  &
   C & \timeform{30'}                 & 0,7   &   10.5  & M & & 44h \\
  120711A                   & 2012-07-11 02:45:07 &  94.703 &  -71.001 &
   B & \timeform{65'}x\timeform{13'}  & 2     &   21.4  & M & 
                                          F[$-14$](15)/L(16)/ &  8h \\
   &&&&&&&&&                        I[$-19$](17)/K[$+49$](18) &     \\
  120908A                   & 2012-09-08 22:35:12 & 230.64  &  -25.79  &
   E & \timeform{28'}x\timeform{19'}  & 4,5   &    6.2  & M & 
                                                F[$-252$](19) & 11h \\
  121025A                   & 2012-10-25 07:46:30 & 248.75  &  +27.73  &
   C & \timeform{17'}                 & 4,5   &    7.0  & M & &  3h \\
  121209A                   & 2012-12-09 21:59:08 & 327.02  &   -7.69  &
   C & \timeform{24'}                 & 4,5   &    9.4  & O & 
                                                  S[$+3$](20) & 111h \\
  121229A                   & 2012-12-29 05:01:09 & 190.10  &  -50.59  &
   C & \timeform{24'}                 & 2,7   &   12.2  & O & 
                                                 S[$-48$](21) &  4h \\
  130102B                   & 2013-01-02 04:42:03 & 309.58  &  -72.38  &
   C & \timeform{12'}                 & 2     &   35.3  & M & 
                         K(22)\footnotemark[$\dagger\dagger$] & 25h \\
  130407A                   & 2013-04-07 23:36:57 & 248.10  &  +10.51  &
   C & \timeform{12'}        & 4,5   &   14.9  & M &       &  4h \\

\hline
\multicolumn{11}{l}{\hbox {\parbox{165mm}{\footnotesize
   \vspace{1ex}
   Notes.  \\
\noindent
\footnotemark[$*$]  
 The center time of the transit in which the burst was observed
\\
\noindent
\footnotemark[$\dagger$] 
 The size of localization error in arcmin. 
 C, B, and E denote the shape of the error; C: circle (radius),
 B: rectangular box (length), and E: ellipsoid (long and short radii), 
 respectively. The systematic errors are not included.
\\
\noindent
\footnotemark[$\ddagger$] 
  Camera-IDs of the cameras that observed the burst.
  There are 12 cameras, camera 0 -- 9, camera A, and camera B.
\\
\noindent
\footnotemark[$\S$] 
 M, O, and H denote that they are found by
\textit{the MAXI Nova-Alert System}, the information of other satellites,
and Human inspection respectively.
\\
\noindent
\footnotemark[$\|$] 
 F : Fermi/GBM, L : Fermi/LAT, S : Swift, I : INTEGRAL, K : Konus-Wind,
 W : Suzaku/WAM.
 The numbers in the square brackets are the trigger (or detection) time 
 of the instrument from the time in the `time' column. 
 The reference numbers are in the parentheses. 
 The references are as follows: 
 (1) \cite{2009GCN..9850....1R},
 (2) \cite{2009GCN..9861....1G},
 (3) \cite{2009GCN..9900....1O},
 (4) \cite{2009GCN..9935....1G},
 (5) \cite{2009GCN..9957....1B},
 (6) \cite{2009GCN..10187...1G},
 (7) \cite{2009GCN..10189...1G},
 (8) \cite{2009GCN..10298...1G},
 (9) \cite{2010GCN..10745...1B},
 (10) \cite{2010GCN..11135...1M},
 (11) \cite{2011GCN..11722...1G},
 (12) \cite{2011GCN..12013...1V},
 (13) \cite{2012GCN..13295...1G},
 (14) \cite{2012GCN..13351...1G},
 (15) \cite{2012GCN..13437...1G},
 (16) \cite{2012GCN..13444...1T},
 (17) \cite{2012GCN..13434...1G},
 (18) \cite{2012GCN..13446...1G},
 (19) \cite{2012GCN..13741...1M},
 (20) \cite{2012GCN..14045...1M},
 (21) \cite{2012GCN..14115...1S},
 (22) \cite{2013GCN..14135...1G}.
\\
\noindent
\footnotemark[$\#$] 
 The time delay in sending GCN or ATel.
\\
\noindent
\footnotemark[$**$] 
 not reported to GCN or ATel
\\
\noindent
\footnotemark[$\dagger\dagger$] 
 The detection time is not reported.

}}}
\end{tabular}
\end{center}
\end{table*}

\begin{table*}[p]
\begin{center}
\caption{X-ray and optical counterparts of MAXI GRBs}
\label{tab:ag}
\begin{tabular}{llD{.}{.}{6}D{.}{.}{6}rrc}
\hline
\hline
  GRB name & Band & 
  \multicolumn{2}{c}{(RA, Dec)} & error & GCN\#\footnotemark[$*$] 
  & redshift (ref.)\footnotemark[$\dagger$] \\
\hline

 090926B & optical  &  46.30808 & -39.00617 & 
     \timeform{0."5} &  9944 & 1.24  (1) \\
 091230  & optical  & 132.91325 & -53.89797 & 
     \timeform{0."5} & 10299 &           \\
 100823A & optical  &  20.70429 &  +5.83511 & 
     \timeform{0."9} & 11148 &           \\
 110213B & optical  &  41.75588 &  +1.14619 & 
                     & 11732 & 1.083 (2) \\
 120510A & X-ray(?)\footnotemark[$\ddagger$] &  44.04666 & +72.88692 & 
     \timeform{4."8} & 13284 &           \\
 120711A & optical  &  94.67850 & -70.99911 & 
                     & 13430 & 1.405 (3) \\
 121025A & X-ray    & 248.38182 & +27.67189 &
     \timeform{3."8} & 13909 &           \\
 121209A & optical  & 326.78733 &  -8.23508 & 
     \timeform{0."5} & 14049 & 2.707 (4) \\
 121229A & optical  & 190.10121 & -50.59430 & 
     \timeform{0."5} & 14117 &           \\

\hline

\multicolumn{7}{l}{\hbox {\parbox{165mm}{\footnotesize
   \vspace{1ex}
   Notes.  \\
\noindent
\footnotemark[$*$] 
 The number of GCN circular which report the position of
 X-ray or optical counterparts.
\\
\noindent
\footnotemark[$\dagger$] 
 The references are 
(1) \cite{2009GCN..9947....1F},
(2) \cite{2011GCN..11736...1C},
(3) \cite{2012GCN..13441...1T},
(4) \cite{2012GCN..14120...1F}.
\\
\noindent
\footnotemark[$\ddagger$] 
 A candidate afterglow was reported, but not confirmed.
}}}
\end{tabular}
\end{center}
\end{table*}

\begin{table*}[htp]
\begin{center}
\caption{A summary of the flux and the spectral hardness}
\label{tab:2}
\begin{tabular}{llccc}
\hline
\hline
  GRB name & 
  \multicolumn{1}{c}{flux\footnotemark[$*$]}
   & hardness\footnotemark[$\dagger$]  
   & $T_d$\footnotemark[$\ddagger$] 
   & other sat.\footnotemark[$\S$] \\
\hline

  090831  &  4.91  $\pm$ 0.14  + 2.68 &  0.47 $\pm$ 0.04 &  41.0 & yes \\
  090926B &  1.68  $\pm$ 0.15   ---   &  1.75 $\pm$ 0.47 &  21.3 & yes \\
  091012  &  0.38  $\pm$ 0.04  + 0.01 &  0.32 $\pm$ 0.09 &  43.4 \\
  091120  &  7.69  $\pm$ 0.23  + 2.54 &  0.46 $\pm$ 0.04 &  29.4 & yes \\
  091201  &  0.47  $\pm$ 0.06  + 0.12 &  0.34 $\pm$ 0.14 &  42.0 \\
  091230  &  0.27  $\pm$ 0.04   ---   &  1.06 $\pm$ 0.38 &  54.7 & yes \\
  100315A &  0.17  $\pm$ 0.03  + 0.27 &  0.25 $\pm$ 0.13 &  26.8 \\
  100327  &  0.26  $\pm$ 0.03  + 0.03 &  0.20 $\pm$ 0.09 &  34.6 \\
  100415A &  0.96  $\pm$ 0.11  + 0.57 &  0.38 $\pm$ 0.13 &  23.4 \\
  100510A &  2.57  $\pm$ 0.19  + 0.75 &  0.54 $\pm$ 0.11 &  28.0 & yes \\
  100616A &  0.37  $\pm$ 0.05  + 0.03 &  0.44 $\pm$ 0.16 &  38.7 \\
  100701A &  0.88  $\pm$ 0.09  + 1.62 &  0.11 $\pm$ 0.04 &  19.8 \\
  100823A &  1.90  $\pm$ 0.10   ---   &  0.24 $\pm$ 0.04 &  22.9 & yes \\
  100911  &  0.30  $\pm$ 0.03  + 0.02 &  0.40 $\pm$ 0.13 &  37.4 \\
  101117A &  0.84  $\pm$ 0.09  + 0.73 &  0.28 $\pm$ 0.09 &  30.0 \\
  101210  &  0.22  $\pm$ 0.03  + 0.03 &  0.48 $\pm$ 0.20 &  33.5 \\
  110213B &  7.58  $\pm$ 0.32   ---   &  0.43 $\pm$ 0.05 &  28.4 & yes \\
  110402  &  0.44  $\pm$ 0.05  + 0.00 &  0.33 $\pm$ 0.11 &  53.0 \\
  110426A &  0.51  $\pm$ 0.06  + 0.79 &  0.21 $\pm$ 0.08 &  17.3 & yes \\
  110916  &  0.27  $\pm$ 0.03  + 0.17 &  0.12 $\pm$ 0.05 &  43.7 \\
  111024A &  2.76  $\pm$ 0.17  + 1.48 &  0.19 $\pm$ 0.04 &  23.6 \\
  120424A &  0.47  $\pm$ 0.05  + 0.08 &  0.22 $\pm$ 0.07 &  32.9 \\
  120510A &  1.53  $\pm$ 0.12   ---   &  0.52 $\pm$ 0.12 &  18.8 & yes \\
  120528B &  1.28  $\pm$ 0.10  + 0.80 &  0.39 $\pm$ 0.09 &  23.5 & yes \\
  120528C &  0.32  $\pm$ 0.05  + 0.04 &  1.17 $\pm$ 0.53 &  36.0 \\
  120614A &  0.96  $\pm$ 0.07  + 0.36 &  0.30 $\pm$ 0.07 &  29.5 \\
  120622A &  0.45  $\pm$ 0.06  + 0.21 &  0.22 $\pm$ 0.10 &  36.0 \\
  120626B &  0.31  $\pm$ 0.04  + 0.20 &  0.27 $\pm$ 0.12 &  23.7 \\
  120711A &  0.99  $\pm$ 0.10   ---   &  0.42 $\pm$ 0.12 &  46.7 & yes \\
  120908A &  0.18  $\pm$ 0.03  + 0.04 &  0.23 $\pm$ 0.12 &  33.1 & yes \\
  121025A &  0.33  $\pm$ 0.06   ---   &  0.22 $\pm$ 0.13 &  14.6 \\
  121209A &  0.52  $\pm$ 0.06   ---   &  0.86 $\pm$ 0.30 &  59.0 & yes \\
  121229A &  0.42  $\pm$ 0.05   ---   &  0.21 $\pm$ 0.08 &  32.7 & yes \\
  130102B &  1.36  $\pm$ 0.11  + 0.49 &  0.35 $\pm$ 0.08 &  44.0 & yes \\
  130407A &  0.41  $\pm$ 0.05  + 0.46 &  0.26 $\pm$ 0.10 &  19.5 \\

\hline
\multicolumn{5}{l}{\hbox {\parbox{165mm}{\footnotesize
   \vspace{1ex}
   Notes.  \\
\noindent
\footnotemark[$*$]
  In the unit of 10$^{-8}$ ergs cm$^{-2}$ s$^{-1}$ in 2--20 keV.
  The first error is statistic error and the second error is
  systematic error due to the uncertainty of the effective area.
  For a burst with accurate position, the systematic error
  is negligible.
\\
\noindent
\footnotemark[$\dagger$]
  Ratio of the photon flux in 8--20 keV to 2--8 keV
\\
\noindent
\footnotemark[$\ddagger$]
  The observed duration, which means the lower limit of the real
  duration of the burst, in the unit of s. 
\\
\noindent
\footnotemark[$\S$]
  The bursts also observed by other satellites are marked
}}}
\end{tabular}
\end{center}
\end{table*}

\begin{table*}[htp]
\begin{center}
\caption{A summary of the spectral parameters and the flux}
\label{tab:spec}
\begin{tabular}{lc cD{.}{.}{9}D{.}{.}{8}D{.}{}{-1}D{.}{.}{8}D{(}{(}{4}}
\hline
\hline
  GRB name & joint & model & 
   \multicolumn{1}{c}{$\alpha$} & \multicolumn{1}{c}{$\beta$}
  & \multicolumn{1}{c}{$E_{\mathrm{peak}}$} 
  & \multicolumn{1}{c}{flux\footnotemark[$*$]}
  & \multicolumn{1}{c}{$\chi^2$(DoF)}
  \\
\hline
  090831  & Fermi
             & PL  & -1.62_{-0.02}^{+0.02} & $---$
   & $---$ & (3.95_{-0.11}^{+0.12}) & 456.23 (261)
  \\
          &  & CPL & -1.27_{-0.04}^{+0.04} & $---$
   & 161._{-20}^{+27} & 4.73_{-0.19}^{+0.15} & 349.23 (260)
  \\
          &  & GRBM\footnotemark[$\dagger$]
                   & -1.03_{-0.08}^{+0.11} & -1.78_{-0.06}^{+0.06}
   & 60._{-15}^{+16} & 5.06_{-0.25}^{+0.10} & 312.60 (259) 
  \\
  090926B\footnotemark[$\ddagger$] & Fermi
             & CPL & 0.44^{+0.14}_{-0.13} & $---$
   & 97._{-6}^{+7} & 1.60_{-0.03}^{+0.04} & 93.26 (83) 
  \\
          &  & GRBM\footnotemark[$\dagger$]
                   & 0.65^{+0.22}_{-0.18} & -2.51^{+0.29}_{-0.49} 
   & 85._{-9}^{+9} & 1.62_{-0.18}^{+0.13} & 83.13 (82) 
  \\
  091120  & Fermi
             & PL  & -1.56$\footnotemark[$\S$]$ & $---$
   & $---$ & (7.60$\footnotemark[$\ddagger$]$) & 417.23 (76)
  \\
          &  & CPL\footnotemark[$\dagger$]
                   & -1.15_{-0.05}^{+0.06} & $---$
   & 90._{-7}^{+9} & 5.72_{-0.22}^{+0.19} & 126.71 (75) 
  \\
          &  & GRBM & -1.14^{-0.06}_{-0.06} & -2.3 $ (fixed)$
   & 83._{-9}^{+10} & 7.23_{-0.22}^{+0.32} & 135.74 (75) 
  \\
  100510A & Fermi
             & PL  & -1.58_{-0.05}^{+0.04} & $---$ 
   & $---$ & 2.54_{-0.24}^{+0.24} & 85.15 (48)
  \\
          &  & CPL\footnotemark[$\dagger$] 
                   & -1.19_{-0.13}^{+0.14} & $---$ 
   & 82._{-16}^{+26} & 2.51_{-0.29}^{+0.24} & 46.87 (47) 
  \\
          &  & GRBM & -1.19_{-0.13}^{+0.16} & -2.3 $ (fixed)$
   & 78._{-20}^{+29} & 2.52_{-0.24}^{+0.44} & 48.27 (47) 
  \\
  100823A & Swift
             & PL  & -2.06_{-0.06}^{+0.06} & $---$ 
   & $---$ & 1.99_{-0.13}^{+0.15} & 53.88 (51)
  \\
          &  & CPL\footnotemark[$\dagger$] 
                   & -1.95_{-0.05}^{+0.15} & $---$ 
   & 8._{-8}^{+10} & 2.02_{-0.17}^{+0.15} & 51.58 (50)
  \\
          &  & GRBM & -1.88_{-0.23}^{+0.36} & -2.3 $ (fixed)$
   & 10._{-9}^{+7} & 2.04_{-2.00}^{+2.13} & 50.95 (50) 
  \\
  110213B & NA
             & PL  & -1.17_{-0.10}^{+0.10} & $---$ 
   & $---$ & 8.12_{-0.56}^{+0.63} & 22.77 (21)
  \\
          &  & CPL\footnotemark[$\dagger$] 
                   & -0.53_{-0.39}^{+0.44} & $---$ 
   & 18._{-5}^{+17} & 8.34_{-5.25}^{+0.17} & 14.49 (20) 
  \\
  111024A & NA
             & PL\footnotemark[$\dagger$]
                   & -2.14_{-0.20}^{+0.18} & $---$ 
   & $---$ & 3.11_{-0.33}^{+0.34} & 18.71 (21)
  \\
          &  & CPL & -1.00 $ (fixed)$         & $---$ 
   & >4.          & 2.95_{-0.53}^{+0.31} & 16.86 (21) 
  \\
  120510A & NA
             & PL\footnotemark[$\dagger$]
                   & -1.30_{-0.27}^{+0.25} & $---$ 
   & $---$ & 1.20_{-0.20}^{+0.24} & 15.43 (18)
  \\
          &  & CPL & -1.00 $ (fixed)$         & $---$ 
   & >12.            & 1.19_{-1.19}^{+0.18} & 14.04 (18) 
  \\

\hline
\multicolumn{7}{l}{\hbox {\parbox{85mm}{\footnotesize
   \vspace{1ex}
   Notes.  \\
\noindent
\footnotemark[$*$] 
  in the unit of 10$^{-8}$ ergs cm$^{-2}$ s$^{-1}$ in 2--20 keV
\\
\noindent
\footnotemark[$\dagger$] 
  The best fit models for each burst are marked.
\\
\noindent
\footnotemark[$\ddagger$] 
  The parameters are from \cite{2011PASJ...63S1035S}.
\\
\noindent
\footnotemark[$\S$] 
  The errors are not available.
\\
}}}
\end{tabular}
\end{center}
\end{table*}

\end{document}